\documentclass[a4paper,10pt,twoside]{cpc-hepnp}

\usepackage{multicol}
\usepackage{graphicx}
\usepackage{booktabs}
\usepackage{amssymb,bm,mathrsfs,bbm,amscd}
\usepackage[tbtags]{amsmath}
\usepackage{lastpage}
\usepackage{CJK}\usepackage{epsfig}

\begin{document}
\begin{CJK*}{GBK}{song}

\title{Effect of tensor force on density dependence of symmetry energy within the BHF Framework}

\author{Pei Wang$^{1,2,3}$, Wei Zuo$^{1,4}$ }
\maketitle

\address{1~(Institute of Modern Physics, Chinese Academy of Sciences, Lanzhou
730000, China)\\
2~(School of Physical Science and Technology, Lanzhou University, Lanzhou 730000, China)\\
3~(Graduate School of Chinese Academy of Sciences, Beijing 100039, China )\\
4~(State Key Laboratory of Theoretical Physics,
 Institute of Theoretical Physics, Chinese Academy of Sciences, Beijing 100190, China)}

\begin{abstract}
The effect of tensor force on the density dependence of nuclear symmetry energy
has been investigated within the framework of the Brueckner-Hartree-Fock approach.
It is shown that the tensor force manifests its effect via the tensor $^3SD_1$ channel.
The density dependence of symmetry energy $E_{sym}$ turns out to be determined essentially by the tensor
force from the $\pi$ meson and $\rho$ meson exchanges via the $^3SD_1$ coupled channel.
Increasing the strength of the tensor component due to the $\rho$-meson exchange tends to
enhance the repulsion of the equation of
state of symmetric nuclear matter and leads to reduction of symmetry energy.
The present results confirm the dominant role played by the tensor force in determining
nuclear symmetry energy and its density dependence within the microscopic BHF framework.
\end{abstract}


\begin{multicols}{2}

\section{Introduction}

Equation of state (EOS) of asymmetric nuclear matter plays
a central role in understanding many physical problems and phenomena in nuclear physics
and nuclear astrophysics, ranging from the structure of rare isotopes
and heavy nuclei~\cite{oyamatsu:1998,avancini:2007,djm:2011,gaidarov:2012}
to the astrophysical phenomena such as supernova
explosions, the structure and cooling properties of neutron stars
\cite{lattimer:1991,zuo:2004a,steiner:2005,fuchs:2006,baldo}.
Nuclear symmetry energy describes the isovector part
of the EOS of asymmetric nuclear matter. To determine the symmetry energy and its
density-dependence in a wide range of density, especially at supra-saturation densities,
 is a new challenge in nuclear
physics and heavy ion physics~\cite{danielewicz:2002,liba:2008}.

Up to now, the density dependence of symmetry energy at low densities below $\sim 1.2\rho_0$ has been
constrained to a certain extent by the experimental observables of heavy ion collisions and some structure information of finite
nuclei, such as isospin-scaling of multifragmentation, isospin transport,
neutron-proton differential collective flow in HIC and neutron skins, pygmy dipole resonances of finite nuclei,
etc~\cite{liba:2008,tsang,chen:2005b,trippa:2008}.
However, the density dependence of symmetry energy at high densities remains poorly known. Different
groups\cite{xiao:2009,feng:2010,russotto:2011} have obtained completely different high-density behaviour of
symmetry energy by comparing the experimental data measured by the FOPI Collaboration
at GSI~\cite{reisdorf:2007} with their calculated results of transport models.

Theoretically, the EOS of asymmetric nuclear matter and symmetry energy
have been investigated extensively by adopting various many-body methods, such as the
Brueckner-Hartree-Fock(BHF)~\cite{bombaci:1991,zuo:1999,zuo:2002,lizh:2006} and
Dirac-BHF~\cite{brockmann:1990,dalen:2005,klahn:2006,sammarruca:2006} approaches,
the in-medium $T$-matrix and Green function
methods~\cite{dewulf:2003,frick:2005,gad:2007,soma:2008,rios:2009},
and the variational approach~\cite{akmal:1998,bordbar:2008}.
Although almost all theoretical approaches are able to reproduce the empirical value of symmetry energy
at the saturation density,
the discrepancy among the predicted density-dependence of symmetry energy at high densities
by adopting different many-body approaches and/or by using different nucleon-nucleon ($NN$) interactions
has been shown to be quite large~\cite{fuchs:2006,Dieperink:2003,lizh:2006,klahn:2006,gogelein:2009}.
In order to clarify the above-mentioned discrepancy among different theoretical predictions, it is desirable to investigate the
microscopic mechanism which controls the high-density behavior of symmetry energy.

The tensor interaction and its effect on finite nuclei and
infinite nuclear matter have been discussed in connection with the evolution of nuclear spectra \cite{wiring:2002}, the
evolution of nuclear shell structure~\cite{to,ms},
the nuclear saturation mechanism~\cite{be,ir}, and the contribution of isovector mesons
to symmetry energy~\cite{cx,fs,iv}. Due to the tensor coupling is
not well determined from experimental data, especially in short-range~\cite{rm1},
there are still a lot of open questions. One of the most important questions is:
what is the role played by  the short-range
tensor force in determining the EOS of asymmetric nuclear matter ?
In Ref.~\cite{iv}, the contributions of various components of $NN$ interaction
to symmetry energy and its slope parameter $L$ at saturation density have been studied
within the BHF framework and it has been shown that the tensor component is decisive for determining
symmetry energy and $L$ around saturation density.
In Ref.~\cite{cx}, the effect of the short-range tensor force due to the $\rho$ meson exchange has been
investigated. In that paper, the tensor force have been added to the Gogny central
force by hand and its effect has been discussed by adjusting the in-medium $\rho$ meson mass
according to the Brown-Rho Scaling.

In the present paper, we shall investigate the effect of tensor force
 on the density dependence of symmetry energy, within the framework of the BHF approach.

\section{Theoretical Approaches}

The present investigation is based on the BHF approach for asymmetric nuclear
matter~\cite{bombaci:1991,zuo:1999}.
Here we simply give a brief review for completeness.
The starting point of the BHF approach is
the reaction $G$-matrix, which satisfies the following isospin
dependent Bethe-Goldstone (BG) equation,
\begin{eqnarray}\label{eq:BG}
&&G(\rho,\beta;\omega)= V_{NN}+
\nonumber\\
 &&V_{NN}\sum\limits_{k_{1}k_{2}}\frac{|k_{1}k_{2}\rangle
Q(k_{1},k_{2})\langle
k_{1}k_{2}|}{\omega-\epsilon(k_{1})-\epsilon(k_{2})}G(\rho,\beta;\omega)
\end{eqnarray}
where $k_i\equiv(\vec k_i,\sigma_1,\tau_i)$ denotes the momentum,
the $z$-component of spin and isospin of nucleon,
respectively. $V_{NN}$ is the realistic $NN$ interaction and $\omega$ is starting energy.
For the realistic $NN$ interaction $V_{NN}$, we adopt the
 $Bonn$-$B$ interaction~\cite{bonn}. The Pauli
operator $Q(k_{1},k_{2})$ prevents two nucleons in intermediate sates from being scattered into their
respective Fermi seas.
The asymmetry parameter $\beta$ is defined as
$\beta=(\rho_{n}-\rho_{p})/\rho$, where $\rho$, $\rho_n$ and $\rho_p$ denote
the total nucleon, neutron and proton
number densities, respectively.
The single-particle (s.p.) energy $\epsilon(k)$ is given by:
$\epsilon(k)={\hbar^{2}k^{2}}/{2m}+U(k)$.
The auxiliary s.p. potential $U(k)$ controls the convergent rate of the
hole-line expansion~\cite{day}.
In the present calculation, we adopt the continuous choice for
the auxiliary potential since it provides a much faster convergence of
the hole-line expansion up to high densities than the gap choice~\cite{baldo:2002}.

In the BHF approximation, the EOS of nuclear matter (i.e., the energy per nucleon of
nuclear matter) is given by~\cite{day}:
\begin{eqnarray}
&&E_A(\rho,\beta)\equiv \frac{E(\rho,\beta)}{A}
=\frac{3}{5}\frac{k_{F}^{2}}{2m} + \nonumber\\
&& \frac{1}{2\rho}Re\sum\limits_{k'\leq
k_{F}}\langle
kk'|G[\rho,\beta;\epsilon(k)+\epsilon(k')]|kk'\rangle_{A} \ .
\end{eqnarray}
One of the main purposes of this paper is to study the density
dependence of symmetry energy which describes the isospin dependent
part of the EOS of asymmetric nuclear matter and is defined generally as:
\begin{equation}
E_{sym} = \frac{1}{2} \left[\frac{\partial ^2 E_A(\rho,\beta)}
{\partial \beta^2}\right]_{\beta=0} \ .
\end{equation}
It has been shown by microscopic investigation~\cite{bombaci:1991,zuo:1999,zuo:2002,gad:2007}
that the energy per nucleon $E_{A}(\rho,\beta)$ of asymmetric nuclear matter
fulfills satisfactorily a linear dependence on $\beta^2$ in the whole asymmetry
range of $0\leq\beta\le1$, indicating that
the EOS of asymmetric nuclear matter can be expressed as:
\begin{equation} \label{eq:Evector}
E_A(\rho,\beta)=E_A(\rho,\beta=0)+E_{sym}(\rho)\beta^{2}
\end{equation}
The above result provides an microscopic support for the empirical
$\beta^2$-law extracted from the nuclear mass table and extended its
validity up to the highest asymmetry.
Accordingly the symmetry energy can be readily obtained from the
difference between the EOS of pure neutron matter and that of symmetric nuclear
matter, i.e.,
\begin{equation}\label{eq:Esym}
E_{sym}=E_A(\rho,\beta=1)-E_A(\rho,\beta=0)
\end{equation}

\section{Results and discussions}

The $Bonn$-$B$ two-body interaction adopted in the present calculation is
an explicit one-boson-exchange potential (OBEP) and
it describes the experimental $NN$ phase shifts with a high precision~\cite{bonn}.
The $Bonn$ potentials content the pseudoscalar $\pi$ and $\eta$ mesons, the
scalar $\sigma$ and $\delta$ mesons, the vector $\rho$ and $\omega$ mesons.
In the OBEP, the tensor components are determined by the competition between the $\pi$
meson and $\rho$ meson exchanges in the isospin singlet ($S=1$, $T=0$) neutron-proton
channel, and they can be written explicitly in configuration space as follows~\cite{bonn}:
\begin{eqnarray}
V_{T}^{\pi}(r)=\frac{1}{12}\frac{f_{\pi}^{2}}{4\pi}m_{\pi}Z(m_{\pi}r)\hat{S}_{12} \ , \\
V_{T}^{\rho}(r)=-\frac{1}{12}\frac{f_{\rho}^{2}}{4\pi}m_{\rho}Z(m_{\rho}r)(1+\frac{f_{\rho}}{g_{\rho}})^{2}\hat{S}_{12} \ ,
\end{eqnarray}
where $\hat{S}_{12}=3\frac{(\sigma_{1}\cdot r)(\sigma_{2}\cdot r)}{r^{2}}-(\sigma_{1}\cdot\sigma_{2})$
is the tensor operator; $f_{\pi}$ and $f_{\rho}$ denote the tensor coupling constants for
the $\pi$ and $\rho$ meson exchanges, respectively; $g_{\rho}$ is the vector coupling constant;
$Z(x)=(m_{\alpha}/m)^{2}(1+3/x+3/x^{2})(e^{-x}/x)$. We notice that a large value of $f_{\rho}/g_{\rho}=6.1$ has
been confirmed and consistently adopted in the OBEP~\cite{bonn,wh1}.
Therefore, the tensor coupling of the $\rho$ meson exchange is much stronger than its vector coupling.
And consequently it may suppress the tensor contribution from the $\pi$ meson exchange at short-range
due to the fact that the mass of $\rho$ meson is much larger than the $\pi$ mass.
The $\pi$ exchange provides a strongly attractive long- and mediate-range tensor component.
A cancelation of the opposite contributions from the $\pi$ meson and $\rho$ meson
exchanges is supposed to generate an mediate-range attractive and
 short-range repulsive tensor force.  One of the most distinctive properties of tensor force is that it couples two-particle
states with different angular momenta of $L=J\pm 1$.

\begin{center}
\includegraphics[width=8cm]{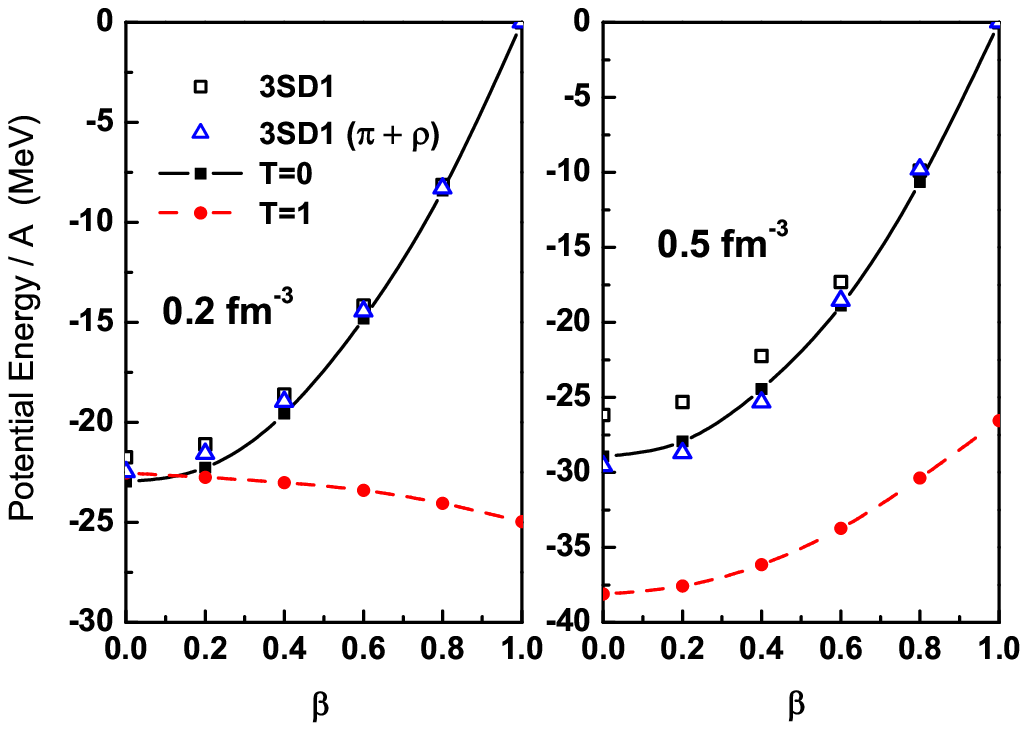}
\figcaption{(Color online) Potential energy per nucleon of asymmetric nuclear matter is
split into the two contributions from the isospin $T=0$ (filled squares) and
$T=1$ (filled circles) channels versus asymmetry parameter $\beta$ for $\rho=0.2fm^{-3}$. The lines
are plotted for guiding eyes. The open squares represent the $^3SD_1$
coupled channel contribution from the full $Bonn$-$B$ interaction. The open triangles denote the $^3SD_1$
channel contribution due to the $\pi$- and $\rho$-exchange parts in the $Bonn$-$B$ interaction.}
\label{fig1}
\end{center}

In order to show the effect of the tensor force on the isospin-dependence of the
EOS of asymmetric nuclear matter, in Fig.~\ref{fig1} we plot
the contributions to the potential energy per nucleon of asymmetric matter at two densities of $\rho=0.2$ and 0.5fm$^{-3}$.
from the isospin singlet $T=0$ channel, the isospin triplet $T=1$ channel and
the $T=0$ $^{3}SD_1$ tensor channel, respectively. In the figure, the results are obtained
by adopting the $Bonn$-$B$ interaction, and the $T=0$ $^3SD_1$ channel
contribution due to the $\pi$ and $\rho$ exchange components in the $Bonn$-$B$ interaction is
also given.
It is seen from Fig.~\ref{fig1} that the contribution of the $T=0$
channel depends strongly on the asymmetry $\beta$ and it increases rapidly
as a function of $\beta$ in the whole asymmetry range $0<\beta<1$. At $\rho=0.2$fm$^{-3}$, it increases from -22.5 MeV to
0 as the asymmetry $\beta$ goes from 0 to 1.
Whereas the asymmetry-dependence of the $T=1$ channel contribution turns out to be
quite weak and it decreases slightly by only about 2.5 MeV from $-22.5$ MeV to $-25$ MeV as the asymmetry
$\beta$ increases from 0 to 1.
The above result is in good agreement with the previous investigation
of Refs.~\cite{bombaci:1991,zuo:1999} within the BHF framework by adopting
the Paris and $AV14$ interactions, respectively. For the high density of $\rho=0.5$fm$^{-3}$, the result remains similar. The $T=0$ channel contribution increases rapidly from -30MeV to 0 as the asymmetry increases from 0 to 1. Whereas, the $T=1$ channel contribution is quite insensitive to the asymmetry parameter $\beta$, and it changes only by about 11MeV from -28MeV to -27MeV. The above results indicate that
the predominant contribution to the isovector part of the potential energy of asymmetric nuclear matter
comes from the isospin singlet $T=0$ channel.
Similar conclusion has also obtained in Ref.~\cite{Dieperink:2003}.
According to Eq.~(\ref{eq:Evector}), the isovector part of the EOS of asymmetric nuclear matter
is completely described by the symmetry energy $E_{sym}$ and therefore the $T=0$ channel contribution
plays a decisive role in determining the symmetry energy and its density dependence.
In Fig.~\ref{fig1}, it is worth noticing that
the corresponding filled squares, empty squares and
empty triangles are almost coincident with one another, which not only indicates that
the $T=0$ channel contribution to the symmetry energy $E_{sym}$ is almost fully comes from
the $^3SD_1$ tensor channel, but also implies that the $T=0$ channel contribution
is provided almost completely by the $\pi$- and $\rho$-exchange
interactions in the tensor $^3SD_1$ channel.
Therefore, we may readily conclude that the potential part of symmetry energy $E_{sym}$ is essentially governed by the tensor
force in $NN$ interaction via the $^3SD_1$ channel.

\begin{center}
\includegraphics[width=8cm]{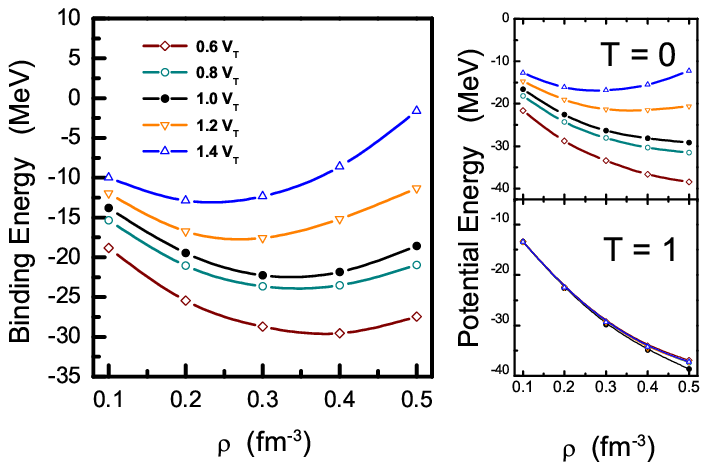}
\figcaption{(Color online) Density dependence of the energy per
nucleon of symmetric nuclear matter obtained by adopting different strengths of
the tensor component due to the $\rho$-meson exchange in the $Bonn$-$B$ interaction.
Left panel: total energy per nucleon. Right panel:
potential parts contributed from the isospin $T=0$ (upper panel) and the $T=1$
(lower panel) channels.} \label{fig2}
\end{center}

The tensor interaction, especially its short-range part,
has not been well determined consistently from the deuteron
properties and/or the nucleon-nucleon scattering data. The tensor
coupling ($D$-state probability $P_{d}\%$) is one of the most uncertain low-energy
parameters, and has estimated to be between $4\%$ and
$8\%$ in various $NN$ potentials~\cite{lizh:2006}.
The tensor force, especially its short-range part, is expect to affect strongly the
high density behavior of symmetry energy~\cite{cx}.
In the following, we shall investigate the short-range tensor force effect on the density-dependence of symmetry energy
by varying the strength of the tensor force due to the $\rho$ meson exchange as follows: $ V_T^{\rho *} = \alpha V_T^{\rho}$,
where $V_T^{\rho}$ is specified as the original tensor component due to the $\rho$-meson exchange in the $Bonn$-$B$ interaction.
By varying the parameter $\alpha$, we may change the strength of the short-range tensor force
from the $\rho$-meson exchange and study its effect.
The calculated results for symmetric nuclear matter are displayed in Fig.~\ref{fig2}.
In the left panel of Fig.~\ref{fig2}, the EOSs of symmetric nuclear matter (i.e., the energy per nucleon of symmetric nuclear matter vs. density)
obtained by adopting various
$\alpha$ values are plotted. As expected, increasing the strength of $\rho$-meson tensor component leads to an overall
increase in the predicted energy per nucleon of symmetric nuclear matter in the whole density region considered here and this effect turns out
to be more pronounced at higher densities.
This is readily understood since the tensor force from the $\rho$-meson exchange
gives a repulsive contribution to the potential energy of nuclear matter. In the right panel, we show the contributions to the
potential energy, respectively from the
isospin $T=0$ channel (upper part) and $T=1$ channel (lower part). It is noticed that the $T=1$ channel contribution is almost independent
of the strength of the $\rho$-meson tensor force, and the variation of
the EOS of nuclear matter with varying the $\rho$-meson tensor force appears to be
determined by the variation of the $T=0$ channel contribution.
The above results indicate that the short-range tensor interaction from the $\rho$-meson exchange play its role essentially
via the $T=0$ partial wave channel.
The $NN$ interaction in the isospin $T=0$ channel describes the neutron-proton ($np$) correlations in nuclear medium.
 Accordingly, our results are consistent with the recent experimental evidence for a strong enhancement of the $np$ short-range correlations
 over the proton-proton($pp$) correlations observed at JLab~\cite{subedi:2008} due to the dominate role played by the short-range
 tensor components of $NN$ interactions in generating the $NN$ correlations~\cite{schiavilla:2007}.
\begin{center}
\includegraphics[width=8cm]{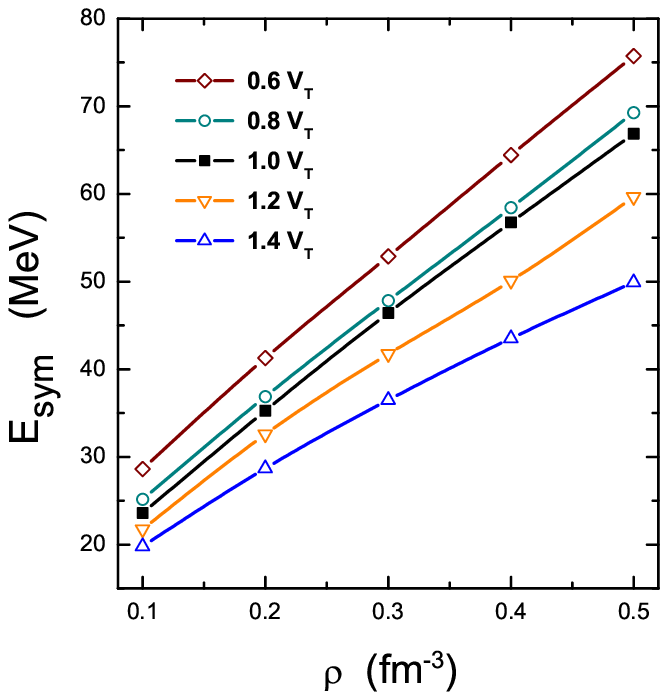}
\figcaption{(Color online) Symmetry energy vs. density, obtained by adopting different strengths of
the $\rho$-meson exchange tensor component in the $Bonn$-$B$ interaction.} \label{fig3}
\end{center}

\par
In Fig.\ref{fig3}, we report the short-range tensor effect on the density dependence
of symmetry energy by varying the strength of the $\rho$-meson
tensor component in the $Bonn$-$B$ interaction.
One may notice that the calculated symmetry energy is rather sensitive to the $\rho$-meson tensor force, especially at high densities
and for large strength parameter $\alpha>1$. The variation of the symmetry energy with varying the tensor component due to the
$\rho$-meson exchange turns out to
be opposite to that of EOS of symmetric nuclear matter, i.e.,
increasing the tensor force of the $\rho$-meson exchange leads to a reduction of symmetry energy and
a softening of the density dependence of symmetry energy. This can be explained easily in terms of Eq.~(\ref{eq:Esym})
and the results in Fig.~\ref{fig2}. According to Eq.~(\ref{eq:Esym}), symmetry energy is determined by the difference between
the energy per nucleon of pure neutron matter and
that of symmetric nuclear matter. From Fig.~\ref{fig2}, it is seen that the short-range $\rho$ exchange tensor force plays its role
almost fully via the $T=0$ channel $np$ correlations which is absent in pure neutron matter.
Therefore, varying the short-range tensor force may lead to opposite variations in symmetry energy and the EOS of symmetric nuclear matter.

\section{Summary}

In summary, we have investigated the effect of tensor force on the isospin dependence of the EOS of asymmetric nuclear matter and
nuclear symmetry energy within the
framework of the BHF approach by adopting the $Bonn$-$B$ interaction.
The $T=0$ channel contribution is shown to depend sensitively on the isospin asymmetry $\beta$, and it
plays a predominate role over the $T=1$ channel contribution
in determining the isospin dependence of the EOS of asymmetric
nuclear matter (i.e., the isovector part of the EOS). The $T=0$ channel contribution stems almost fully
from the $^3SD_1$ tensor channel.
The tensor force manifests its effect via the $^3SD_1$ coupled channel, and
the $T=0$ channel contribution turns out to be almost completely provided by the tensor component in $NN$ interaction
via the $^3SD_1$ coupled channel.
The contributions to the EOS of symmetric nuclear matter and symmetry energy
from the short-range tensor component due to the $\rho$-meson exchange in $NN$ interaction
have been calculated.
The $T=1$ channel contribution to the EOS is almost independent of the strength of the $\rho$-meson exchange tensor component.
Whereas, the $T=0$ channel contribution is shown to be affected strongly by the $\rho$-meson exchange tensor component.
Increasing the tensor component due to the $\rho$-meson exchange in $NN$ interaction tends to
 enhance the repulsion of the EOS of symmetric nuclear matter, and may leads to a reduction and softening of symmetry energy.
 The present results confirm the crucial role played by the tensor force in determining
the isospin dependence of the EOS of asymmetric nuclear matter and the density dependence of nuclear symmetry energy.

\section*{Acknowledgments}
We would like to thank U. Lombardo, B.A. Li, L.G. Cao, G. C. Yong
for valuable discussions. The work is supported by
the 973 Program of China (No. 2007CB815004),
 the National Natural Science Foundation of China
(11175219), the Knowledge
Innovation Project(KJCX2-EW-N01) of Chinese Academy of Sciences.

\end{multicols}
\end{CJK*}
\end{document}